# Deviation of the statistical fluctuation in heterogeneous anomalous diffusion

Yuichi Itto

*Science Division, Center for General Education, Aichi Institute of Technology,
Aichi 470-0392, Japan*

**Abstract.** The exponent of anomalous diffusion of virus in cytoplasm of a living cell is experimentally known to fluctuate depending on localized areas of the cytoplasm, indicating heterogeneity of diffusion. In a recent paper (Itto, 2012), a maximum-entropy-principle approach has been developed in order to propose an *Ansatz* for the statistical distribution of such exponent fluctuations. Based on this approach, here the deviation of the statistical distribution of the fluctuations from the proposed one is studied from the viewpoint of Einstein's theory of fluctuations (of the thermodynamic quantities). This may present a step toward understanding the statistical property of the deviation. It is shown in a certain class of small deviations that the deviation obeys the multivariate Gaussian distribution.

There exists a remarkable phenomenon, which is exotic from the traditional perspective of the physics of diffusion. Such a phenomenon has experimentally been observed in the infection pathway of adeno-associated viruses in living *HeLa* cells by the use of the technique of real-time single-molecule imaging [1-3]. (Here, the adeno-associated virus is a small virus particle, whereas the *HeLa* cell is a line of human epithelial cells.) The experiments show that, in cytoplasm of the cell, the virus, which is labeled with fluorescent dye molecule, exhibits stochastic motions in both the free form and the form being contained in the endosome (i.e., a spherical vesicle). There, the mean square displacement in stochastic motion, which is denoted here by $\overline{x^2}$, is evaluated based on analysis of the trajectories of the fluorescent viruses. Then, $\overline{x^2}$ scales for large elapsed time, $t$, as

$$\overline{x^2} \sim t^\alpha. \tag{1}$$

The experimental results show not only normal diffusion, leading to $\alpha = 1$, but also subdiffusion, corresponding to $0 < \alpha < 1$. Remarkably, subdiffusion of the virus exhibits a novel feature [2] that the exponent, $\alpha$, fluctuates depending on localized areas of the cytoplasm: $\alpha \in (0.5, 0.9)$. This may not be due to the forms of existence of the virus (i.e., the free or endosomal forms) [2] and, thus, highlights its *heterogeneity*, in marked contrast to anomalous diffusion widely discussed over the years, for example, in Refs. [4-10] (see Ref. [11] for a recent review).

In recent works [12,13], a kinetic theory has been developed in order to describe the infection pathway of the virus over the cytoplasm by generalizing fractional kinetics [14] modeling anomalous diffusion in a unified way. There, the statistical distribution of the fluctuations of $\alpha$ over the cytoplasm plays a central role. According to the experiment [2], 113 trajectories of the viruses are analyzed. The result of the analysis is as follows. In the form in Eq. (1), 53 trajectories show $\alpha = 1$, and 51 exhibit $\alpha$ varying between 0.5 and 0.9. On the other hand, the mean square displacement in 9 trajectories has a parabolic form for the elapsed time, indicating diffusion with drift. The number of these trajectories is seen to be less compared to those in the case of normal diffusion and subdiffusion (a relevant discussion can be found in Ref. [3]). Therefore, the result in 9 trajectories has been neglected in the discussions in Refs. [12,13]. (Following Refs. [12,13], the result is also not taken into account in the present discussion.) From the above result on 104 trajectories, as the statistical property of the fluctuations of $\alpha$, it is considered [12,13] that normal diffusion is often to be realized, whereas subdiffusion with the exponent near $\alpha = 0$ may seldom be the case. This consideration is also motivated by the property [2] that the virus tends to reach the nucleus of the cell. In addition, the exponents found in both the free and endosomal forms are supposed to be different from each other only slightly. Consequently, as an *Ansatz* for the statistical distribution of the fluctuations, the following Poisson-like distribution, which is expressed here in the case of discrete values of the exponent for convenience in our later discussion, has been proposed:

$$P_{\alpha_i} \propto e^{\lambda \alpha_i} \quad (i = 1, 2, \ldots, A), \tag{2}$$

where $A$ is the total number of different values of the exponent, $\alpha_i$ is the $i$ th value of the exponent, and $\lambda$ is a positive constant.

We here wish to mention the following. The distribution in Eq. (2) has property of monotonic increase with respect to the exponent. On the other hand, as can be seen in Fig. 7.44 in Ref. [1], in which the experimental data of the frequency of the exponent in the case of subdiffusion is presented, the weights of the exponents decrease at $\alpha = 0.8$. This indicates that the statistical distribution of fluctuations to be suggested by the weights has no property of monotonic increase in the whole range of $\alpha$ described there, although it is seen to have such a property in the range near $\alpha = 1$. Therefore, one might think that Eq. (2) does not well explain the distribution based on the weights. Regarding this point, however, it seems necessary to clarify if each trajectory is taken from each of different localized areas in the cytoplasm. Then, the distribution in Eq. (2) is supposed to describe the statistical fluctuation on a large spatial scale in the cytoplasm, not limited to the localized areas studied in the experiments. Accordingly, it seems fair to say that this distribution is yet to be carefully examined with further information on the fluctuations. In Refs. [12,13], it has been shown that it is in fact possible to theoretically derive the distribution in Eq. (2) (which is continuous there) in a consistent manner (see the later discussion), supporting its realization. Therefore, we suppose that this is the distribution of relevance, here.

Now, a fundamental premise in the discussions in Refs. [12,13] is that the time scale of variation of exponent fluctuations is much larger than that of stochastic motion of the virus in each localized area: that is, the exponent slowly varies. The exponent is then assumed to be approximately constant during the motion of the virus over the cytoplasm.

Here, a natural question arises. If this assumption is relaxed, then the statistical distribution of the fluctuations to be observed on a long time scale may deviate from that in Eq. (2) due to variation of the fluctuations, in general, although the deviation seems to be small. Accordingly, the question is how the behavior of such a deviation can be determined.

In this short note, we wish to answer this question for a certain class of deviations. For it, from the viewpoint of Einstein's theory of fluctuations [15-17], we examine the approach proposed in Refs. [12,13], where it is shown that the Poisson-like distribution of exponent fluctuations can be derived by the maximum entropy principle. Since the deviation should not be so large due to slow variation of the fluctuations, we consider, as the statistical distribution of the fluctuations to be observed, a distribution that can deviate from the Poisson-like distribution in Eq. (2) only slightly. Specifically, we focus our attention on a class of deviations in the following situation: the expectation values of the exponent with respect to both this distribution and that in Eq. (2) are equal to each other. Such a class is of physical interest in accordance with the above approach. Then, we show that the deviation in this class obeys the multivariate Gaussian distribution. The present discussion can be seen as a step toward understanding the statistical property of the deviation.

First, we recapitulate the approach in Refs. [12,13] to derive the distribution in Eq. (2). The situation considered there is as follows. The cytoplasm, over which the exponent slowly varies locally, is regarded as a medium for stochastic motion of the virus in the free form as well as the endosomal one. Then, this medium is imaginarily divided into many small blocks with a set of different values of the exponent, $\{\alpha_i\}_{i=1,2,...,A}$. A block is identified with a localized area of the cytoplasm. The medium can be thought of as a collection formed by constructing those blocks. It is noted that there is no information available about how the exponent locally distributes over the cytoplasm. Accordingly, all of possible distinct collections, each of which is different from each other in terms of the local fluctuations, are identical to the medium at the statistical level of the fluctuations. In this situation, the distribution in Eq. (2) can be obtained by the maximum entropy principle. Let $N$ and $n_{\alpha_i}$ be the total number of blocks in the medium and the number of blocks with the exponent $\alpha_i$ in the medium, respectively. The following quantity is then evaluated as the entropy:

$$S = \frac{\ln G}{N}, \qquad (3)$$

where $G$ is the total number of distinct collections. Recalling that the cytoplasm is regarded as a medium being imaginarily divided into a lot of blocks, each of which is characterized by each value of the exponent in Eq. (1), $G$ is estimated as follows. It is understood from Eq. (1) that the exponent in each block of the medium is obtained by average based on the trajectory of virus, in which trajectories in other blocks are not included. So, it is supposed that the exponent in a given block is determined independently of those in other blocks. That is, the blocks are supposed to be independent each other in terms of the exponent. From these considerations, $G$ is given by $G = N! / \prod_{i=1}^{A} n_{\alpha_i}!$. Under the assumption that $N$ and $n_{\alpha_i}$'s are large since the medium is composed of many blocks, the entropy in Eq. (3) is expressed, by using the Stirling approximation [i.e., $\ln(M!) \cong M \ln M - M$ for large $M$], in the form of the Shannon entropy:

$$S \cong S[P] = -\sum_{i=1}^{A} P_{\alpha_i} \ln P_{\alpha_i}, \qquad (4)$$

where $P_{\alpha_i} = n_{\alpha_i} / N$ is the probability of finding the exponent $\alpha_i$ in a given block of the medium. The same symbol $P_{\alpha_i}$ as that in Eq. (2) is used for this probability, but it will not cause any confusion. The situation under consideration is that information is only available about the statistical property of the local fluctuations. So, $S[P]$ is maximized under the constraints on the expectation value of the exponent, $\sum_{i=1}^{A} \alpha_i P_{\alpha_i} = \bar{\alpha},$ as well as the normalization condition, $\sum_{i=1}^{A} P_{\alpha_i} = 1$:

$\delta_P \left\{ S[P] - \kappa \left( \sum_{i=1}^{A} P_{\alpha_i} - 1 \right) + \lambda \left( \sum_{i=1}^{A} \alpha_i P_{\alpha_i} - \bar{\alpha} \right) \right\} = 0,$ where $\kappa$ and $\lambda$ are the

Lagrange multipliers associated with the normalization condition and the expectation value, respectively, and $\delta_P$ stands for the variation with respect to $P_{\alpha_i}$. Note that the condition, $P_1 > P_{\alpha_{\min}}$ with $\alpha_{\min}$ being $\alpha_{\min} \equiv \min\{\alpha_i\}_{i=1,2,\ldots,A}$ in the range $0 \leq \alpha_{\min} < 1$, which turns out to require $\lambda$ to be a positive Lagrange multiplier, has been imposed. This is because of the following. Assuming that diffusion of the virus is subdiffusion as well as normal diffusion, it is considered that the virus in any block tends to reach the nucleus, although there might exist blocks with $\alpha = 0$ in Eq. (1). This tendency is seen to become strong in the case of normal diffusion rather than subdiffusion, since due to the time-dependence of $\overline{x^2}$ in Eq. (1), the virus seems to move relatively fast to neighboring blocks in the former compared to the latter. Accordingly, taking the above-mentioned tendency into account, the number of blocks with the value $\alpha = 1$ in Eq. (1) is naturally supposed to be larger than that of blocks with the value $\alpha_{\min}$ (which is taken to be $\alpha_{\min} = 0$ in Ref. [13]), here, suggesting the above condition. Consequently, the resulting stationary solution is given by $\hat{P}_{\alpha_i} \propto e^{\lambda \alpha_i}$, which has, in fact, the form in Eq. (2).

Before proceeding, it may be of interest to examine how large $N$ and $n_{\alpha_i}$'s are. Below, we present a brief discussion about this issue.

In Ref. [2], the cell is considered as a sphere with the radius 5 μm. So, we regard the nucleus of the cell as a sphere with the radius $2\,\mu\text{m}$, here. Correspondingly, in this case, the volumes of the cell and nucleus are estimated as $524\,\mu\text{m}^3$ and $34\,\mu\text{m}^3$, respectively, indicating that the volume of the cytoplasm is given by $490\,\mu\text{m}^3$.

Now, we treat a given local block in the medium as a cubic one having, as the length of its side, the value of $\sqrt{\overline{x^2}}$ at large elapsed time, here. Accordingly, for the experimental data presented in Fig. 3 (G) in Ref. [2] (see also Fig. 3 in Ref. [3]), the volumes of the cubic block at the elapsed time $0.30\,\text{s}$ are estimated as follows: $2.2\,\mu\text{m}^3$ for the free form in the case of normal diffusion, but $0.54\,\mu\text{m}^3$ and $0.24\,\mu\text{m}^3$ for the endosomal forms in the case of normal diffusion and subdiffusion with the exponent $\alpha = 0.6$, respectively. If the total number of blocks with each of these volumes in the medium is, for example, $50$ and there are no additional blocks with $\alpha = 1$ as well as $\alpha = 0.6$, then it follows that $n_1 = 100$, $n_{0.6} = 50$ and $N > 150$. Now, in this case, the corresponding total volumes of the blocks with the volumes estimated above are respectively given by $110\,\mu\text{m}^3$, $27\,\mu\text{m}^3$, and $12\,\mu\text{m}^3$, showing that the volume of the cytoplasm mentioned above is large compared to the sum of these volumes. Therefore, this observation implies that $n_1$ and $n_{0.6}$ are actually larger than the above values and the numbers of blocks with other exponents in the medium are necessarily large in order for the volume of the medium to be equal to that of the cytoplasm. These considerations also seem to suggest that the assumption on $N$ and $n_{\alpha_i}$'s mentioned earlier is appropriate, (supporting the use of the Stirling approximation).

Next, let us address ourselves to studying the deviation of the statistical distribution of the fluctuations to be observed on a long time scale from the Poisson-like distribution $\{P_{\alpha_i}\}_{i=1,2,...,A}$ in Eq. (2) based on the maximum-entropy-principle approach.

A key point is that the total number of blocks in the medium, $N$, appearing in Eq. (3) is supposed to be large. This is seen to allow us to discuss the deviation from the viewpoint of Einstein's theory of fluctuations [15] [see Eq. (6) below]. To see this, let $G_{\max}$ be the maximum value of $G$, for which the entropy in Eq. (3) has its maximum value: $S_{\max} = (\ln G_{\max})/N$. $S_{\max}$ is evaluated by $S[P]$ with $\{P_{\alpha_i}\}_{i=1,2,...,A}$ in Eq. (2), here. Then, suppose that $G_{\max}$ is decreased to a certain value, $G'$, which is given, for example, by $G' = G_{\max}/e^k$ with $k$ being a positive constant. Correspondingly, $S_{\max}$ is also decreased to $S' = (\ln G')/N = S_{\max} - k/N$. Therefore, if $k/N$ is so small that it is negligible compared to $S_{\max}$, then $S'$ is approximately equal to $S_{\max}$, implying that the statistical fluctuation corresponding to $G'$ can be realized. In this sense, we assume in what follows that the statistical distribution of the fluctuations to be observed, which deviates from the Poisson-like distribution only slightly, is given by that of the fluctuations corresponding to a value of $G$ smaller than $G_{max}$.

With this assumption, we consider not only the set of $n_{\alpha_i}$'s leading to $G_{max}$ but also other accessible sets in the above sense, each of which is different from each other in terms of the values of $n_{\alpha_i}$'s. Recalling the maximum-entropy-principle approach to the distribution in Eq. (2), it is clear that these accessible sets fulfill the same conditions as those to be satisfied by the set associated with $G_{max}$. In particular, we suppose for the accessible sets that the condition $P_1 > P_{\alpha_{min}}$ on the probabilities appearing in Eq. (4) is automatically fulfilled since the deviation is small, here. (This point will be mentioned again in terms of the deviation in the later discussion.) Denoting the value of $G$ of the $n$th set in all of the sets by $G^{(n)}$, we describe the probability of finding the medium in the state with the statistical fluctuation associated with the $n$th set by

$$W^{(n)} = \frac{G^{(n)}}{\sum_m G^{(m)}}, \qquad (5)$$

where the summation is taken over all of the sets. Later, we shall see how $W^{(n)}$ in Eq. (5) depends on the deviation of the statistical fluctuation from the Poisson-like one.

Now, we immediately see that $W^{(n)}$ has the following form:

$$W^{(n)} \propto e^{NS^{(n)}}, \qquad (6)$$

where $S^{(n)}$ is the entropy associated with $G^{(n)}$ given by $S^{(n)} = (\ln G^{(n)})/N$. Equation (6) has an analogy with Einstein's theory of fluctuations [15-17]: the reversal of Boltzmann's relation, $W \propto e^{S}$, where $S$ is the thermodynamic entropy, (provided that the Boltzmann constant is set equal to unity). This notation for the thermodynamic entropy should not be mixed with that for the entropy in Eq. (3).

Let us proceed to evaluate $W^{(n)}$ in Eq. (6). We describe the statistical distribution of the fluctuations associated with the $n$ th set as follows:

$$P_{\alpha_i}^{(n)} = P_{\alpha_i} + \Delta P_{\alpha_i}^{(n)} \quad (i = 1, 2, \ldots, A). \qquad (7)$$

Here, $P_{\alpha_i}$ is given in Eq. (2) and $\Delta P_{\alpha_i}^{(n)}$ stands for the deviation of the statistical distribution from $P_{\alpha_i}$ in Eq. (2), which is assumed to be small. Clearly, the normalization condition, $\sum_{i=1}^{A} P_{\alpha_i}^{(n)} = 1$, should be fulfilled. Then, the expectation values of the exponent with respect to both $\{P_{\alpha_i}^{(n)}\}_{i=1,2,\ldots,A}$ and $\{P_{\alpha_i}\}_{i=1,2,\ldots,A}$ are equal to each other, i.e., $\sum_{i=1}^{A} \alpha_i P_{\alpha_i}^{(n)} = \sum_{i=1}^{A} \alpha_i P_{\alpha_i} = \bar{\alpha}$, as mentioned earlier. For the deviation, $\{\Delta P_{\alpha_i}^{(n)}\}_{i=1,2,\ldots,A}$, they, respectively, lead to the following two conditions:

$$\sum_{i=1}^{A} \Delta P_{\alpha_i}^{(n)} = 0, \qquad \sum_{i=1}^{A} \alpha_i \, \Delta P_{\alpha_i}^{(n)} = 0. \qquad (8)$$

Although $P_1^{(n)} > P_{\alpha_{\min}}^{(n)}$ yields another condition on the deviation, but we suppose that the deviation automatically fulfills such a condition due to its smallness, here. It should be noticed that for the present approach to be applicable, the number of different values of the exponent, $A$, has to satisfy $A > 2$, since otherwise the deviation identically vanishes. According to the experimental data [1,2], this is seen to be fulfilled.

Now, from Eq. (4), the entropy $S^{(n)}$ can be evaluated by $S[P^{(n)}]$ with $\{P_{\alpha_i}^{(n)}\}_{i=1,2,\ldots,A}$. Accordingly, expanding $S[P^{(n)}]$ up to the second order of $\Delta P_{\alpha_i}^{(n)}$'s, using the conditions in Eq. (8) [and noting that $S_{\max} \cong S[P]$ with $\{P_{\alpha_i}\}_{i=1,2,\ldots,A}$ in Eq. (2)], we find

$$S^{(n)} \cong S_{\max} - \frac{1}{2} \sum_{i=1}^{A} \frac{1}{P_{\alpha_i}} \left( \Delta P_{\alpha_i}^{(n)} \right)^2. \qquad (9)$$

Substituting Eq. (9) into Eq. (6), we obtain the probability of the following form:

$$W^{(n)} \sim W_{\max} \exp\left[ -\frac{N}{2} \sum_{i=1}^{A} \frac{1}{P_{\alpha_i}} \left(\Delta P_{\alpha_i}^{(n)}\right)^2 \right] \qquad (10)$$

with $W_{\max} \equiv G_{\max} / \sum_m G^{(m)}$ being the probability of finding the medium in the state with the Poisson-like fluctuation. Using the conditions in Eq. (8), one can express, for example, each of $\Delta P_{\alpha_{A-1}}^{(n)}$ and $\Delta P_{\alpha_A}^{(n)}$ in terms of the deviation, $\{\Delta P_{\alpha_i}^{(n)}\}_{i=1,2,\ldots,A-2}$. Therefore, $W^{(n)}$ in Eq. (10) also turns out to have the following form:

$$W^{(n)} \sim W_{\max} \exp\left[ -\frac{N}{2} \sum_{i,j=1}^{A-2} h_{ij} \Delta P_{\alpha_i}^{(n)} \Delta P_{\alpha_j}^{(n)} \right] \qquad (11)$$

with the positive-definite symmetric matrix, $H = (h_{ij})$, having the following element:

$$h_{ij} = \frac{1}{P_{\alpha_i}} \delta_{ij} + \frac{1}{P_{\alpha_{A-1}}} \frac{(\alpha_i - \alpha_A)(\alpha_j - \alpha_A)}{(\alpha_{A-1} - \alpha_A)^2} + \frac{1}{P_{\alpha_A}} \frac{(\alpha_i - \alpha_{A-1})(\alpha_j - \alpha_{A-1})}{(\alpha_{A-1} - \alpha_A)^2}. \qquad (12)$$

Equation (10) [or equivalently (11)] explicitly shows that the deviation obeys the multivariate Gaussian distribution.

In conclusion, we have studied the deviation of the statistical distribution of exponent fluctuations in heterogeneous anomalous diffusion of an adeno-associated virus in cytoplasm of a living *HeLa* cell. To develop the discussion, we have examined the maximum-entropy-principle approach to the statistical distribution from the viewpoint of Einstein's theory of fluctuations. Considering a statistical distribution that can slightly deviate from the Poisson-like distribution, we have concentrated on a certain class of small deviations, in which the expectation values of the exponent with respect to both this distribution and the Poisson-like one are equal to each other. Then, we have shown that the deviation in this class follows the multivariate Gaussian distribution. The present approach may provide new insight into the study of the heterogeneity of anomalous diffusion.